\documentclass[english,twocolumn,showpacs,preprintnumbers,amsmath,amssymb,superscriptaddress,prl]{revtex4}
\usepackage{mathptmx}
\usepackage[T1]{fontenc}
\usepackage[latin9]{inputenc}
\usepackage{textcomp}
\usepackage{graphicx}

\makeatletter
\@ifundefined{textcolor}{}
{%
 \definecolor{BLACK}{gray}{0}
 \definecolor{WHITE}{gray}{1}
 \definecolor{RED}{rgb}{1,0,0}
 \definecolor{GREEN}{rgb}{0,1,0}
 \definecolor{BLUE}{rgb}{0,0,1}
 \definecolor{CYAN}{cmyk}{1,0,0,0}
 \definecolor{MAGENTA}{cmyk}{0,1,0,0}
 \definecolor{YELLOW}{cmyk}{0,0,1,0}
}

\makeatother

\usepackage{babel}
\begin{document}

\title{Arbitrary Bending Plasmonic Light Waves }

\author{Itai Epstein}

\author{Ady Arie}

\email{Corresponding author: ady@eng.tau.ac.il}

\affiliation{Department of Physical Electronics, Fleischman Faculty of Engineering,
Tel Aviv University, Tel Aviv 69978, Israel}
\begin{abstract}
We demonstrate the generation of self-accelerating surface plasmon
beams along arbitrary caustic curvatures. These plasmonic beams are
excited by free-space beams through a two-dimensional binary plasmonic
phase mask, which provides the missing momentum between the two beams
in the direction of propagation, and sets the required phase for the
plasmonic beam in the transverse direction. We examine the cases of
paraxial and non-paraxial curvatures and show that this highly versatile
scheme can be designed to produce arbitrary plasmonic self-accelerating
beams. Several different plasmonic beams, which accelerate along polynomial
and exponential trajectories, are demonstrated both numerically and
experimentally, with a direct measurement of the plasmonic light intensity
using a near-field-scanning-optical-microscope. 
\end{abstract}
\maketitle
Surface-plasmon-polaritons (SPPs) are surface electromagnetic waves
that are coupled to electron waves, which propagate at the interface
between a dielectric and a metallic medium \cite{Maier_SPP_book}.
The ability to control and guide plasmonic light waves opens exciting
new possibilities in photonics and electronics \cite{Atwater HA.,shalaev}.
Specifically, nanoscale on-chip technologies such as surface plasmon
circuitry \cite{ebbesen}, sub-wavelength optical devices \cite{barnes,bozhevlny}
and nanoscale electro-optics \cite{ozbay}, as well as new applications
in biology and chemistry such as bio-sensing, optical trapping and
micro-manipulation at the nanoscale \cite{quidant} have attracted
great interest in recent years.

In the last several years, new types of plasmonic beams have been
realized having unique properties. These beams can be ``non-spreading''
- i.e. preserve their spatial shape with propagation, as well as ``self-accelerating''
- i.e. propagate along curved trajectories. For example, the plasmonic
Cosine-Gauss beam \cite{cappaso} is a non-spreading beam which propagates
along a straight trajectory, whereas the plasmonic Airy beam \cite{salandrino,minovitch,Li,peng}
is non-spreading and propagates along a parabolic trajectory. The
latter, is the only self-accelerating plasmonic beam demonstrated
until now and is restricted to a parabolic trajectory. In this paper,
we address the question of whether it is possible to create self-accelerating
surface plasmon beams that propagate along arbitrary curved trajectories.

The Airy function is an exact solution of the paraxial Helmholtz equation,
or equivalently, of the Schrodinger equation for a free particle \cite{Berry}
which carries infinite energy. An actual Airy beam, however, carries
finite energy and is obtained by truncating the infinitely long tail
of the Airy function by using an exponential or Gaussian window \cite{siviloglou}.
The truncated Airy beam preserves its shape and self-accelerates,
but only over a finite distance. Recently it was shown \cite{Greenfield,frohely}
that free-space non-spreading beams, propagating along arbitrary convex
trajectories over finite distances, can be realized. However, the
question still remains whether this concept, demonstrated so far only
with free-space beams, can be used for the case of surface plasmons
waves. If so, several fundamental challenges, owing to the plasmonic
nature of the waves, should be addressed.

First, coupling a surface plasmon wave from a free-space wave requires
a compensation for the missing momentum between the two wave-vectors,
as the plasmonic wave-vector $k_{spp}$ is always greater than that
of the free-space wave $k_{0}$ \cite{Maier_SPP_book}. Second, owing
to the limited propagation length of surface plasmons and the limited
measurement range of characterization tools such as near-field-scanning-optical-microscopes
(NSOM), a significant acceleration is required over a fairly short
propagation distance (typically <100 microns), meaning that the paraxial
approximation would not be valid. Third, while planar phase plates
readily provide a well defined phase pattern for a free-space beam
at the entrance plane, the surface plasmon is excited over a finite
propagation distance and therefore it's phase cannot be simply defined
at a specific one-dimensional plane. Fourth and last, dynamic tools
for controlling the wavefront of free-space beams, like spatial-light-modulators
(SLM), do not exist for surface plasmons. Despite these fundamental
and practical challenges, we demonstrate here a robust method to excite
self-accelerating surface plasmons propagating along arbitrary caustic
trajectories. Specifically, we generate surface plasmons which propagate
along several different polynomial and exponential trajectories.

To address the obstacles mentioned above, we introduce a two-dimensional
binary plasmonic phase mask, which is analytically described by the
following equation:

\begin{equation}
t(z,y)=\frac{h_{0}}{2}\left\{ 1+sign\left[cos\left(\frac{2\pi}{\Lambda}z+\phi_{i}(y)\right)\right]\right\} 
\end{equation}

The mask is modulated periodically in the direction of propagation
$z$ with period $\Lambda$. The resulting grating can compensate
for the missing momentum between the two wave vectors, $k_{spp}$
and $k_{0}$, by one it's wave-vectors $k_{G}=2\pi m/\Lambda$, where
$m$ is an integer and $h_{0}$ is the grating's height. Hence the
momentum conservation equation in the direction of propagation is:
\begin{equation}
k_{spp}=k_{in}+k_{G}
\end{equation}
where $k_{in}=k_{0}sin\alpha$ and $\alpha$ is the illumination angle.
Furthermore, in order to excite an SPP that will follow a caustic
trajectory, the desired initial phase $\phi_{i}(y)\equiv\phi(y,z=0)$
is encoded in the transverse direction $y$. We emphasize that in
contrast to planar phase plates for free-space beams, which operate
only in the transverse direction, this binary plasmonic phase mask
operates both in the propagation direction and in the transverse direction. 

\begin{figure}
\includegraphics[width=1\columnwidth]{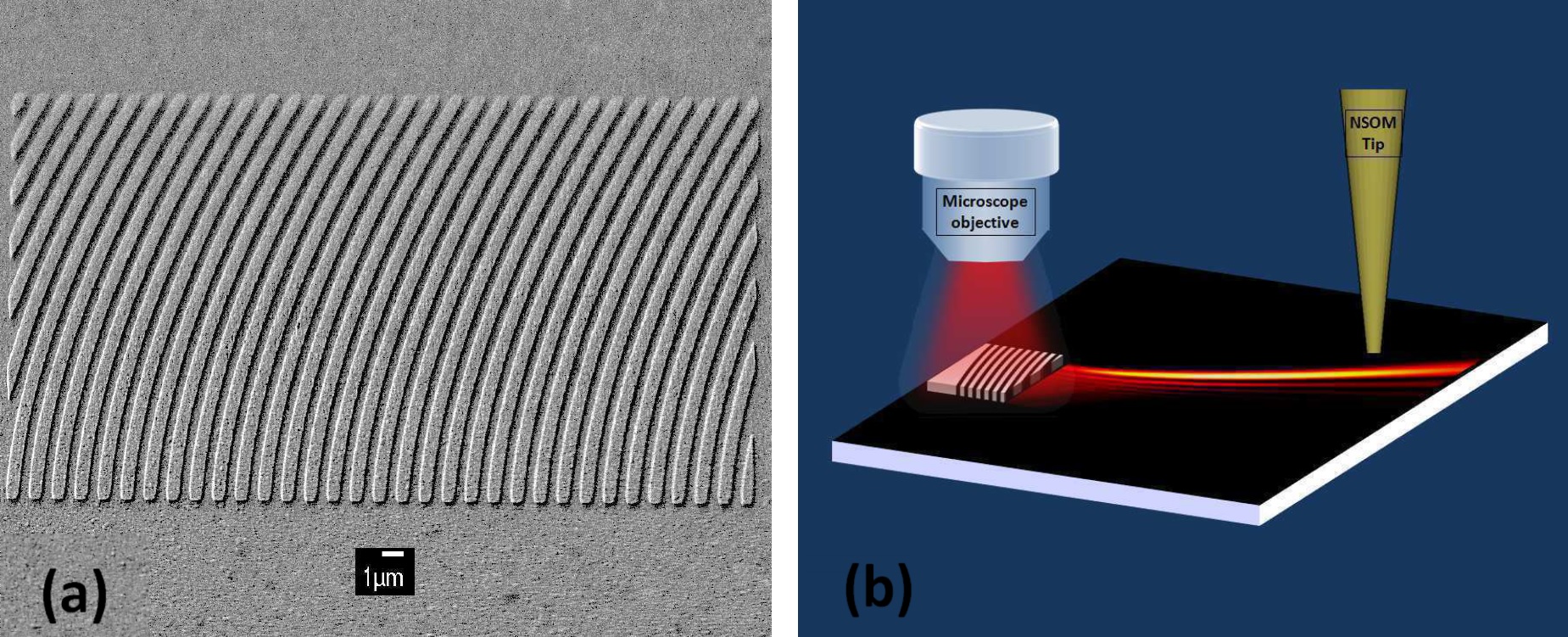}

\caption{(a) SEM image of the fabricated binary plasmonic phase mask for the
case $y=a_{1}z^{1.5}$. (b) Experimental setup.}
\end{figure}

It is now required to derive the initial phase $\phi_{i}(y)$ corresponding
to the desired analytical curve $y=f(z)$. The derivation is based
on the principle that the caustic curve $f(z)$ can be constructed
by multiple geometrical rays, which are tangent to the curve itself
. In our case, the transverse modulation of the phase generates these
geometrical rays at angles $\theta(y)$ with respect to the $z$ axis,
where $\frac{d\phi_{_{i}}(y)}{dy}=ksin[\theta(y)]$ \cite{frohely}.
This sets the relation between the angle $\theta(y)$ and the caustic
trajectory $f(z)$ to be: 

\begin{equation}
\frac{d\phi_{i}(y)}{dy}=\frac{kf'(z)}{\sqrt{1+[f'(z]^{2}}}
\end{equation}
where $f'(z)=df(z)/dz$. Under the paraxial approximation, this relation
can be further simplified \cite{Greenfield} thereby enabling to obtain
analytic expressions for the transverse phase of various trajectories. 

First we examine the case of trajectories under the paraxial approximation
and follow the procedure given by \cite{Greenfield} to derive the
phase for the following curves: $y=a_{1}z^{1.5}$, $y=a_{2}z^{2}$,
$y=a_{3}z^{3}$ and $y=b_{1},exp(q_{1}z)$, where $a_{1}=34.1494\cdot10^{-3},a_{2}=2.4691\cdot10^{-3},a_{3}=2.7435\cdot10^{-5},b_{1}=5\cdot10^{-7},q_{1}=40\cdot10^{-3}$
are arbitrary chosen constants (in micrometer units). For example,
for the trajectory of $y=a_{1}z^{1.5}$ the required transverse phase
is $\phi_{i}(y)=\frac{9}{4}(\frac{a_{1}}{2})^{\frac{2}{3}}y^{\frac{4}{3}}$.

The fabrication of the plasmonic phase masks was done by evaporating
$200nm$ of silver on a BK7 glass substrate, followed by electron-beam
lithography of the mask pattern on PMMA (polymethyl methacrylate).
A $50nm$ silver layer was evaporated above the PMMA, followed by
a lift-off process. The final device, shown in Fig.1(a), consisted
of a $50nm$ thick binary plasmonic phase mask, on top of a $200nm$
layer of silver. The experimental setup, shown in Fig.1(b), was composed
of a fiber-coupled diode laser ($\lambda=1.064\mu m$), focused on
the mask by a microscope objective lens, and a Nanonics MultiView
2000\texttrademark{} NSOM system that was used to measure the plasmonic
light intensity. All the plasmonic phase masks were designed for free-space
illumination at normal incidence $\alpha=0$, at which the surface
plasmon wavelength is equal to the period of modulation $\Lambda$
and the $m=1$ order of the grating satisfies Eq.(2).

\begin{figure}
\includegraphics[width=1\columnwidth]{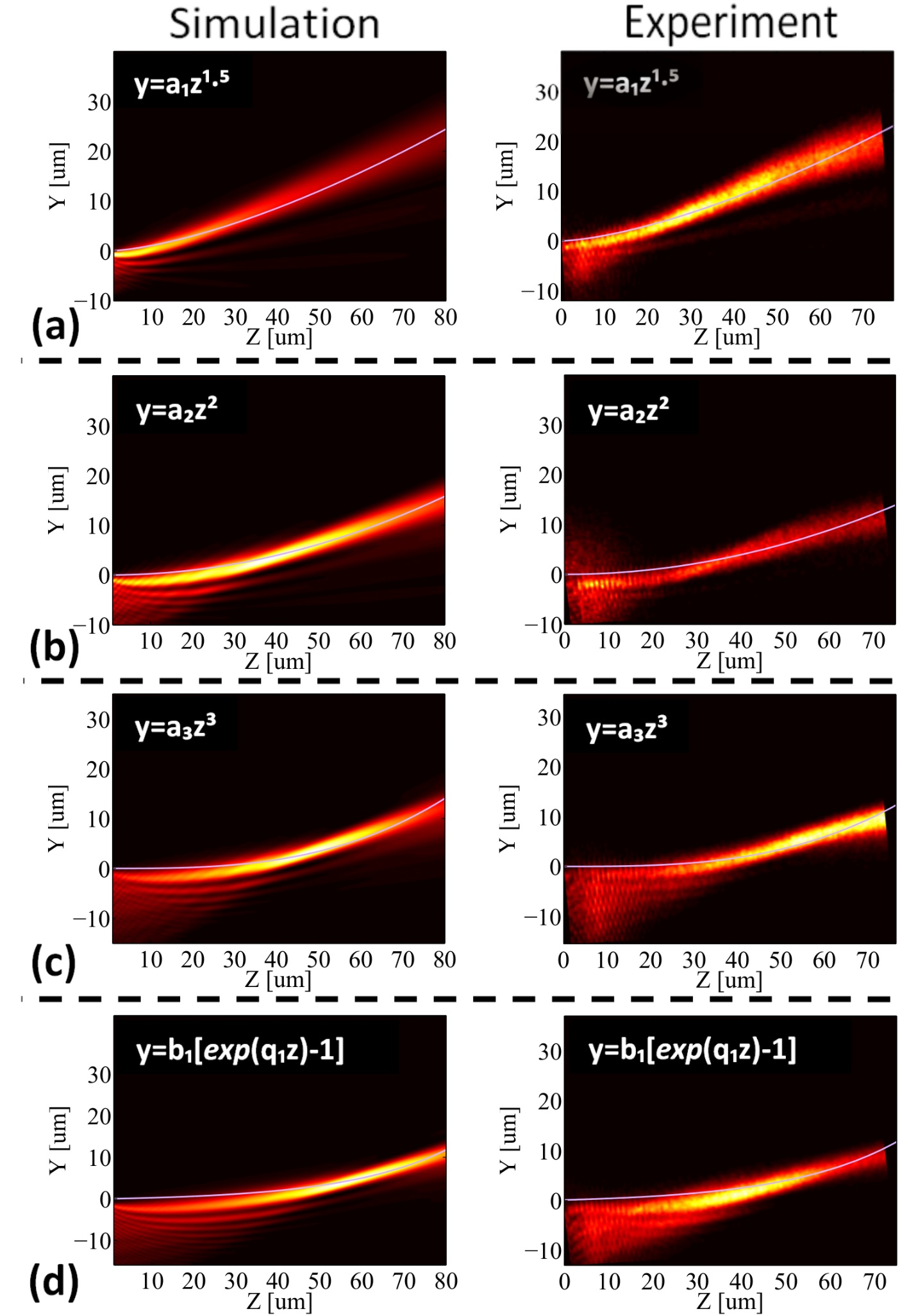}

\caption{Numerical simulations and experimental results for the following analytical
curves derived under the paraxial approximation: (a) $y=a_{1}z^{1.5}$,
(b) $y=a_{2}z^{2}$, (c) $y=a_{3}z^{3}$ and (d) $y=b_{1}exp(q_{1}z)$.
The purple solid curves in all figures depicts the target analytical
curve $y=f(z)$. }
\end{figure}

\begin{figure}[!t]
\includegraphics[width=1\columnwidth]{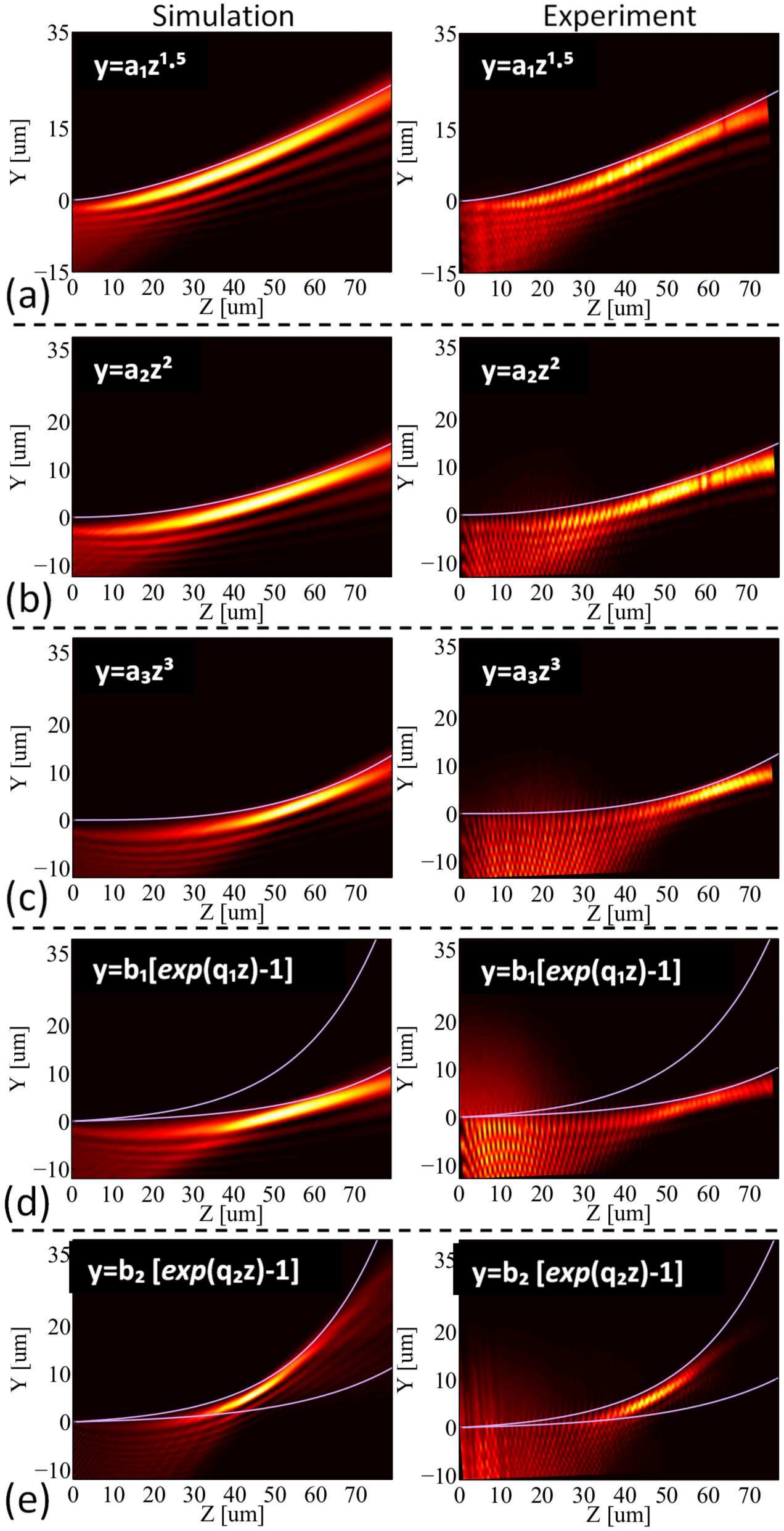}

\caption{(a)-(e) numerical simulations and experimental results for the same
analytical curves presented in Fig.2, only this time derived for non-paraxial
curvatures. }
\end{figure}

To simulate the intensity distribution of the generated plasmon waves
we realized a numerical calculation based on the two-dimensional Green's
function of the Helmholtz equation. The experimental results and the
numerical simulations for the different curves, under the paraxial
approximation, are presented in Fig.2, and it is clearly seen that
the experimental results are in good agreement with the simulations.
We note however that both numerical simulations and experimental results
exhibit diffraction and deviation from the target analytical curve,
depicted by the purple solid curve, in all of the examined trajectories.
We believe that this is a manifestation of the paraxial approximation
used to derive $\phi_{i}(y)$. In our experiment, all curves were
designed to exhibit the desired acceleration within $80\mu m$, corresponding
to the scanning range limit of the NSOM system. This rapid acceleration
is actually already beyond the paraxial limit, making this method
applicable only to weaker accelerations, spanned on larger length
scales. We also note that all figures were rotated between $3-5$
degrees as to align the coordinates of the target analytical curves
with those of the scanning NSOM system. We relate the periodic intensity
variations, transverse to the direction of propagation, appearing
in the experimental results within the individual beam lobes, to interference
between the plasmonic beam and the back-reflected free-space beam
recorded by the NSOM, thus they depend on the phase of the plasmonic
beam \cite{Yin}.

\begin{figure}[!t]
\includegraphics[width=1\columnwidth]{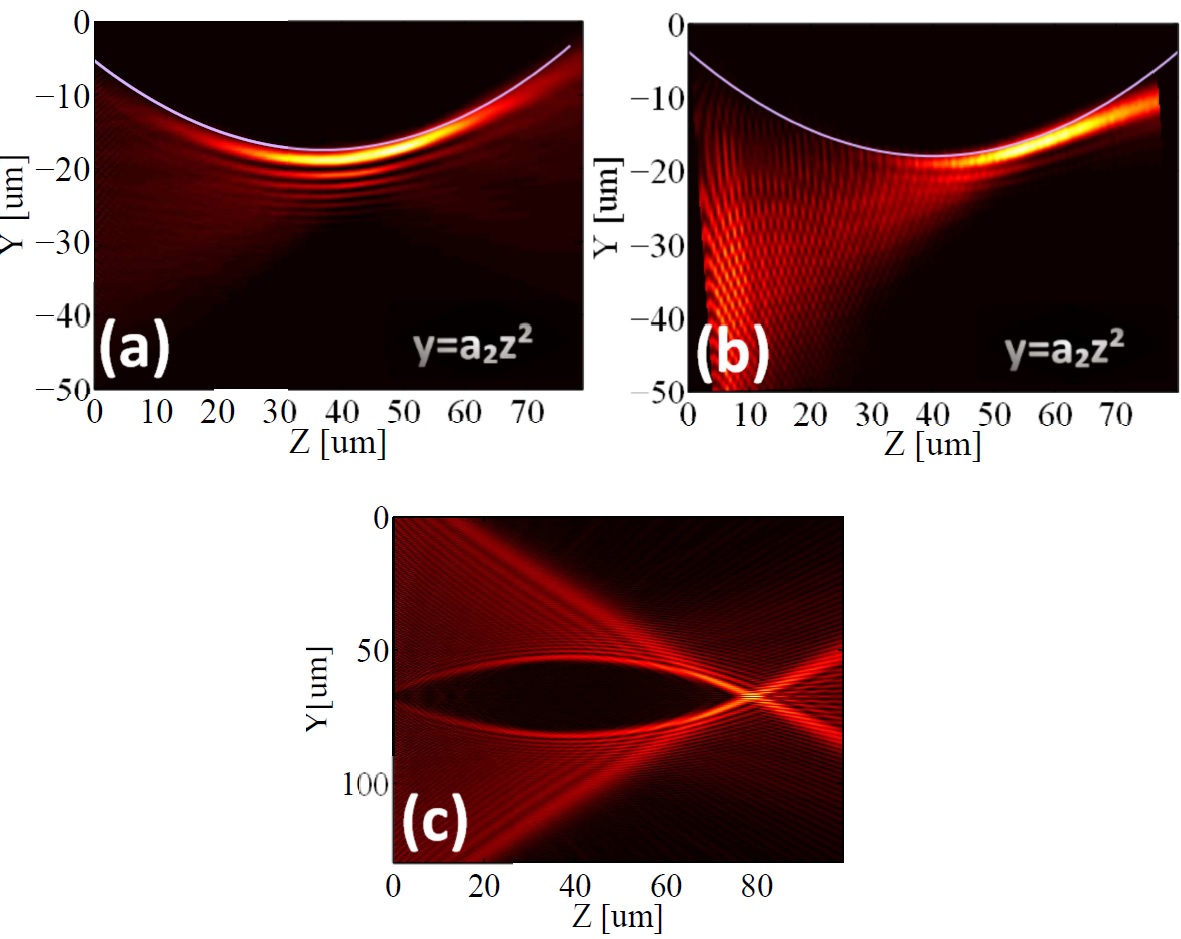}

\caption{(a) numerical simulations and (b) experimental results of a non-monotonic
parabolic accelerating plasmonic beam. (c) Numerical simulation of
a plasmonic bottle-beam. }
\end{figure}

We therefore examined next the non-paraxial case by following the
procedure given by \cite{frohely} to derive $\phi_{i}(y)$ using
Eq.3, for the same analytical curves. In this case Eq.3 can be solved
numerically in order to derive $\phi_{i}(y)$. Both the numerical
simulations and the experimental results for the non-paraxial case
are presented in Fig.3. Once again, the agreement between the simulations
and the measurements is clearly seen, but this time the beam closely
follows the target analytical curve $y=f(z)$. The small deviations
of the beam from the curve towards the end of the scan range are attributed
to the Gaussian shaped illumination of the the free-space beam, which
results in lower amplitude at the edges, and can therefore be reduced
by illuminating the mask uniformly. Fig.3(d),(e) show the results
for two exponential trajectories, each with different constants, and
both trajectories are shown in both figures for comparison. In some
of the measurements, a round halo surrounding the plasmonic beam can
be observed owing to back-reflections of the free-space beam.

We can therefore conclude that for rapid accelerations on such short
distances the non-paraxial method is indeed more suitable. Furthermore,
this method is not limited only to monotonically increasing caustic
curves. In order to illustrate the flexibility of the method, we show
in Fig.4(a),(b) (simulation and measurement, respectively), a parabolic
trajectory but the initial transverse phase is set at a plane which
is $40\mu m$ to the left of the parabola's vertex. The trajectory
therefore bends downwards at first and then, after the vertex point,
rises upwards. The mask that generated the beam is presented in Fig.5(a).
This concept can be further extended by a mask that generate two mirror-imaged
trajectories recombined after a certain distance. This would be a
two-dimensional ``area-caustic'' or ``area-bottle'' beam version
of the ``volume-caustic'' beam demonstrated by \cite{frohely},
or optical bottle-beam demonstrated by \cite{Chremmos} in free-space,
and may enable to trap particles to the dark area confined by the
caustic curves \cite{quidant}. As an example, a simulation of such
a mask realizing a plasmonic bottle-beam is presented in Fig.4(c).

\begin{figure}
\includegraphics[width=1\columnwidth]{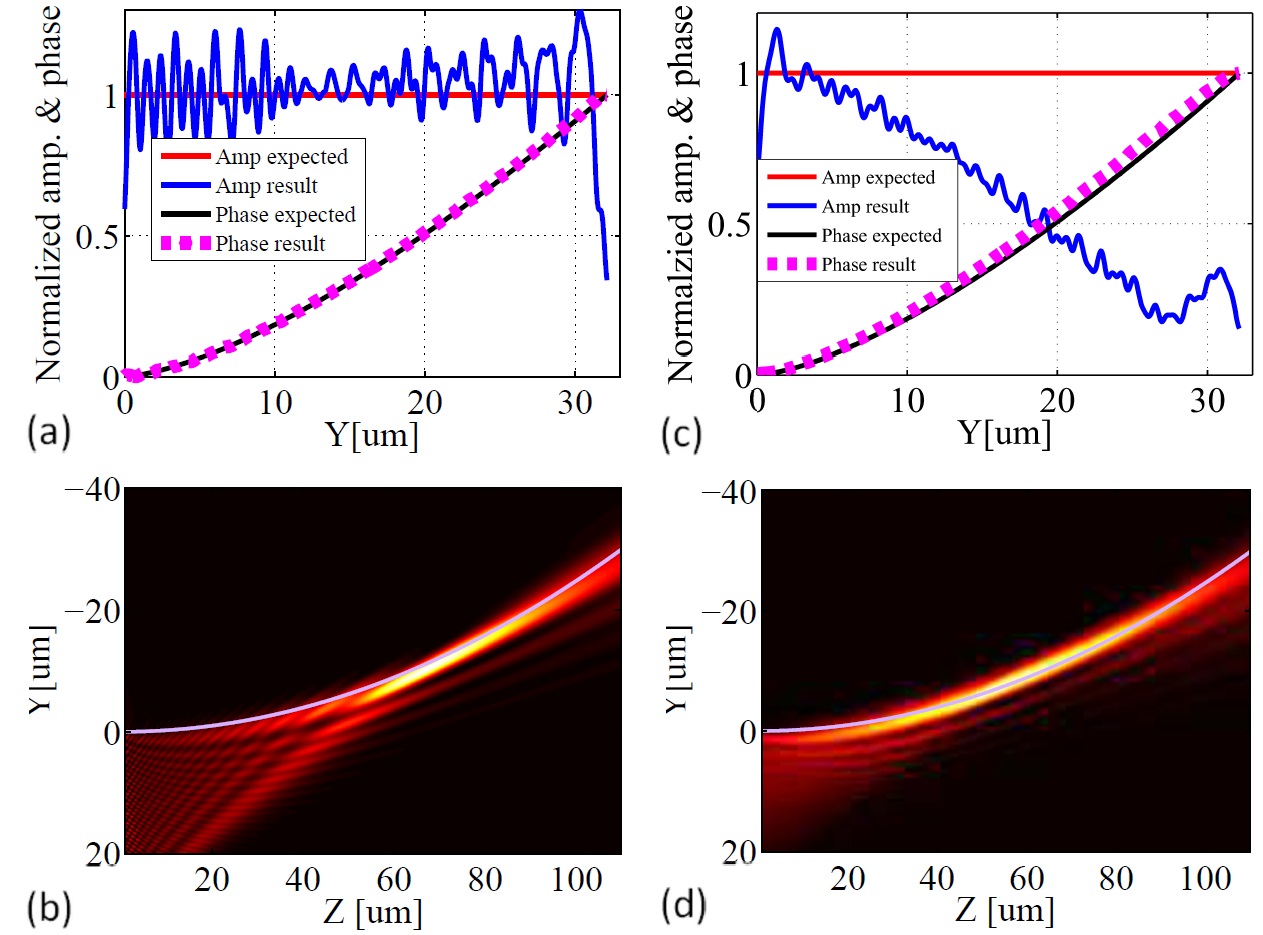}

\caption{Expected and resulting amplitude (red and blue curves, respectively)
and accumulated phase (black and purple dashed curves, respectively)
emanating from (a) single-cycle mask and (c) nine-cycles mask, for
a $y=a_{2}z^{2}$ curve. Plasmonic beam intensity emanating from (b)
single-cycle, and (d) nine-cycles masks. }
\end{figure}

Next we analyze the design considerations of the optimal plasmonic
phase mask, with the main design parameter being the number of periodic
cycles within the mask. For an ideal phase-only mask, the field emanating
from the mask should have a uniform amplitude and the required one-dimensional
phase $\phi_{i}(y)$. While this is easily obtained for free-space
beams, reflected from a planar phase mask or SLM, in the case of the
plasmonic phase mask this ideal state can be realized by applying
the mask with a single cycle. Fig.5(a) showes simulations of the resulting
amplitude and accumulated phase emanating from the single-cycle mask,
for the case of a $y=a_{2}z^{2}$. Unfortunately, coupling SPPs via
a single-cycle grating is not an efficient process and results in
a weak SPP intensity. However, adding additional periodic cycles to
the mask results in two different effects - increasing the coupling
efficiency on one hand, but changing the resulting amplitude and phase
from their target values on the other hand. These effects are presented
in Fig.5(c) for a mask with nine periodic cycles. It can be seen that
the amplitude is not uniform anymore and the accumulated phase exhibits
a small increase at larger $y$ values. The change in the accumulated
phase leads to a deviation from the target analytical curve and the
change in amplitude yields a change in the intensity distribution
of the beam. These are presented in Fig.5(b),(d).

In order to understand the origin of these effects, let's examine
Eq.(1). The underlying assumption of that equation is that the propagation
direction $z$ and transverse direction $y$ can be separated. However,
since the plasmonic beam is accelerating, the transverse modulation
also has an effect on the propagation axis. This can be seen in Fig.6(a)
- the thickness of the grooves in the $y$ direction at the bottom
of the mask is smaller than those at the top. This means that the
light density being coupled at that area is smaller, thereby leading
to the unwanted variation in the amplitude of the generated surface
plasmon. To resolve this issue we have considered a transformation
of coordinates coinciding with those of the accelerating surface plasmon
frame $\widetilde{y}=y+f(z)$. This however, yielded only a small
improvement. We therefore found that the optimal solution is to limit
the number of periodic cycles together with the transformation coordinates.
This solution was already implemented in the masks of the non-paraxial
beams and a SEM image of the mask for the case of $y=a_{1}z^{1.5}$
is presented in Fig.6(b). We do emphasize, however, that implementing
a single-cycle plasmonic phase mask can circumvent these problems
with the cost of reduced coupling efficiency.

\begin{figure}
\includegraphics[width=1\columnwidth]{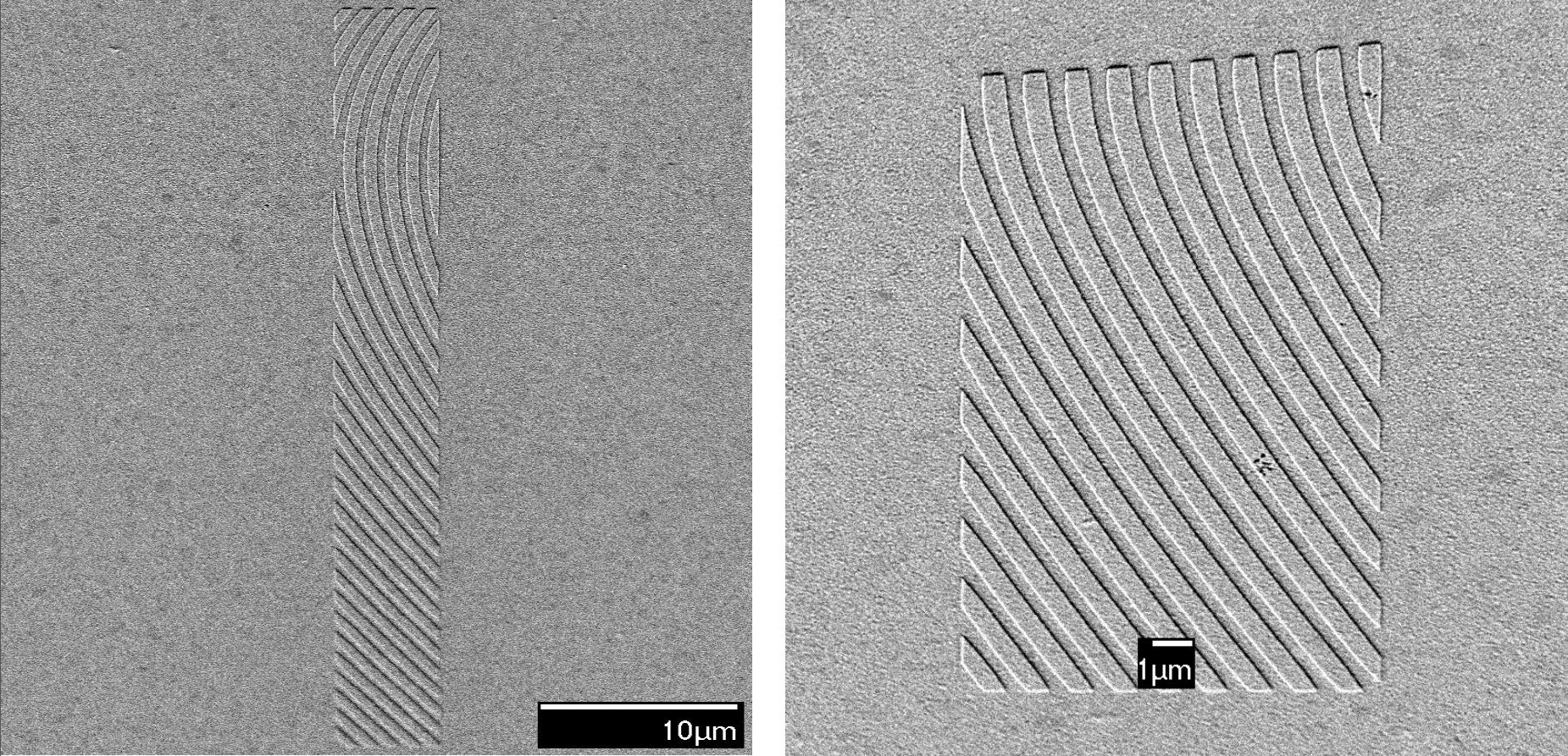}

\caption{SEM scan of the fabricated plasmonic phase masks for the case of (a)
non-monotonic parabolic and (b) $y=a_{1}z^{1.5}$ trajectories..}
\end{figure}

To conclude, we have demonstrated numerically and experimentally the
generation of self-accelerating plasmonic light beams that propagate
along arbitrary caustic trajectories. We examined the cases of paraxial
and non-paraxial caustics and found the latter to be more suitable
for the case of rapidly accelerating plasmonic light beams. We discussed
the design limitations of the plasmonic phase mask and found the crucial
parameter to be the number of periodic cycles used in the mask. We
believe that this demonstration of arbitrary self-accelerating surface
plasmon waves will enable new exciting possibilities in photonics
and electronics at the nanoscale. For example, these beams can enable
the trapping and guiding of micro-particles along the arbitrary curves,
or to circumvent an obstacle by designing a bypassing caustic. Moreover,
we expect the method will be used for shaping the caustics of other
types of waves, such as surface acoustic waves \cite{slobodnik},
ground radio waves, electron waves \cite{Noa_nature}, etc.

\end{document}